\documentclass[12pt]{article}
\usepackage{epsf,amsmath, amsfonts}
\usepackage{epsfig,subfigure}
\usepackage{multirow}
\usepackage{color,graphicx}
\usepackage{dcolumn}
\usepackage{bm}
\usepackage{epsfig,latexsym}
\usepackage{lineno}
\usepackage{chemarr,dashrule}
\newcommand{\ben}{\begin{eqnarray}}
\newcommand{\een}{\end{eqnarray}}
\newcommand{\beq}{\begin{equation}}

\newcommand{\eeq}{\end{equation}}
\newcommand{\bea}{\begin{array}}
\newcommand{\eea}{\end{array}}
\newcommand{\bef}{\begin{figure}}
\newcommand{\eef}{\end{figure}}

\def\Eus{E_{us}}

\long\def\drop#1{}

\begin{document}

\title{\large \bf Determining equilibrium osmolarity in Poly(ethylene glycol) / Chondrotin sulfate gels mimicking articular cartilage} 
\author{ {\bf S. Sircar$^{a,e}$, E. Aisenbrey$^b$, S. J. Bryant$^{b, c, d}$ \& D. M. Bortz$^{a, c}$}\\{$^a$ \small Department of Applied Mathematics, University of Colorado, Boulder, CO 80309} \\{\small $^b$ Department of Chemical \& Biological Engg., University of Colorado, Boulder, CO 80309} \\{\small $^c$ BioFrontiers Institute, University of Colorado, Boulder, CO 80309}\\{\small $^d$ Material Science and Engineering Program, University of Colorado, Boulder, CO 80309}\\{\small $^e$ Corresponding Author (email: sarthok.sircar@colorado.edu)}}
\maketitle

\begin{abstract}
We present an experimentally guided, multi-phase, multi-species polyelectrolyte gel model to make quantitative predictions on the electro-chemical properties of articular cartilage. The mixture theory consists of two different types of polymers: Poly(ethylene gylcol) (PEG), Chondrotin sulfate (ChS), water (acting as solvent) and several different ions: H$^+$, Na$^+$, Cl$^-$. The polymer chains have covalent cross-links modeled using Doi rubber elasticity theory. Numerical studies on polymer volume fraction and net osmolarity (difference in the solute concentration across the gel) show the interplay between ionic bath concentrations, pH, polymer mass in the solvent and the average charge per monomer; governing the equilibrium swelled / de-swelled state of the gel. We conclude that swelling is aided due to a higher average charge per monomer (or a higher percentage of charged ChS component of the polymer), low solute concentration in the bath, a high pH or a low cross-link fraction. However, the swelling-deswelling transitions could be continuous or discontinuous depending upon the relative influence of the various competing forces.
\end{abstract}


\section{Introduction} \label{sec:intro}
Articular cartilage is a polyelectrolyte biogel that forms a thin tissue layer lining the articulating ends of all diarthrodial joints in the body and contribute to the lubrication mechanisms in the joint \cite{Mow1997}. The water phase of cartilage constitutes on average from 60-80 \% of the total weight for normal tissue while the dominant structural components of the solid matrix are the collagen molecules and proteoglycans (PGs) \cite{Mow1993}. The collagen fibrils are densely packed polymer strands with a high resistance against fluid flow \cite{Armstrong1982}, which helps in retaining the shape of the cartilage when compressed \cite{Maroudas1979}. PG are macromolecules consisting of a protein core to which are attached short chains of negatively charged glycosaminoglycans (GAGs) \cite{Setton1993}. The primary GAGs associated with PGs in cartilage are chondroitin 4-sulfate, chondroitin 6-sulfate and keratan sulfate \cite{Maroudas1979}. 


An increase in cartilage tissue hydration, governed by the density and the nature of fixed charges on the PGs as well as the density of the mobile counter ions in the interstitial fluid are the earliest signs of articular cartilage degeneration during osteoarthritis \cite{Armstrong1982, Maroudas1979}. For this reason the development of methods to quantify cartilage swelling have been of great interest for many years. Each PG-associated negative charge, on the polymer chain, requires a mobile counter-ion (e.g., Na$^+$) dissolved within the interstitial fluid in the gel to maintain electro-neutrality within the gel \cite{Maroudas1991}. This gives rise to an imbalance of mobile ions across the gel interface. The excess of mobile ions colligatively yields a swelling pressure, known as the osmotic pressure \cite{Maroudas1968}, while the swelling pressure that is associated with the fixed charges on the polymer chain is known as the Donnan pressure \cite{Donnan1924}. The swelling pressure in articular cartilage is balanced by tensile forces generated in the collagen network \cite{Muir1970}. Thus, changes in this internal swelling pressure, arising from altered ion concentrations of the external bath, or changes in the fixed charges on the polymer network will result in changes in tissue dimensions and hydration.

Previous work on polyelectrolyte gels builds on  the classic work of Tanaka on hydrogels \cite{Tanaka1980}, and includes effects of osmotic pressure arising from the charge-induced Donnan equilibrium \cite{English1996,Wolgemuth2004}, the Poisson-Boltzmann equation and Biot's theory of poroelasticity \cite{De2002}, as well as phenomenological modifications of Flory-Higgins theory to capture some effects of multi-ionic cations \cite{Kokufuta2005,Liverpool2006}.   Furthermore, in previous gel swelling literature, the Flory interaction parameter was either chosen to be a function of temperature alone or of temperature and polymer volume fraction \cite{Wolgemuth2004, Hirotsu1990}.

The purpose of this paper is to present a comprehensive model of polyelectrolyte gel swelling/deswelling mechanism, mediated by the dissolved ions in the solvent as well as covalent cross-links within the polymer chain. To calibrate the outcome of this model, in vitro set-up of cross-linked copolymerized poly(ethylene glycol) (PEG) chondrotin sulfate (ChS) and polymethacrylate gels were developed. The kinetic chains associated with polymethacrylate group represents less than one percent of the total hydrogel volume and therefore they are not considered in the model. PEG was chosen since it can be functionalized to enable cross-linking to form a 3-D matrix, a system that promotes the deposition of proteoglycans and collagen molecules when cartilage cells are encapsulated and emulates the mechanical strength, load bearing capabilities and resilience of cartilage tissue \cite{Bryant2003}. The ChS component serves two purposes: to mimic the biochemical environment of cartilage (since it is the main component of proteoglycans) and to introduce fixed negative charges into the network \cite{Bryant2004}. 

In the next section, we present the details of this model, including the equation of motion and interface conditions (\S \ref{EoM}), chemical potentials (\S \ref{CP}) and the chemistry of the binding reaction at quasi-equilibrium conditions (\S \ref{Chem}). In \S \ref{sec:MM}, we provide the details of the gel swelling experimental set-up, while the results pertaining to the equilibrium configuration of these ionic gels under different electro-chemical stimuli are presented in \S \ref{sec:results}. We conclude with a brief discussion of the implication of these results and the focus of our future directions.

\section{Multi-species, multi-phase cartilage-gel model} \label{sec:model}
We view the PEG-gel with negatively charged ChS strands in ionic solution, as a multi-component material, including solvent particles, polymers, and several ionic species. The polymer is assumed to be made up of two types of monomers which are either uncharged (i.e PEG segment of the polymer) or the charged units (which is the ChS segment of the polymer), denoted as M, and each of which carries a double negative charge (or M$^{2-}$). The positively charged ions in the solvent are Hydrogen (H$^+$) and Sodium (Na$^+$). The negatively charged ions could include Hydronium (OH$^-$) and Chloride (Cl$^-$). Because the negatively charged ions are assumed to be not involved in any binding reactions with the gel, acting only as counterions to positive charges, we identify these ions by the name Chloride. The binding reactions of the positively charged ions with the monomers are: \vskip -10pt
 \ben
&&(a) \hspace{0.25cm} {\rm M^{2-}}+{\rm H^+} \xrightleftharpoons[\text{k$_{-h}$}]{\text{k$_h$}} {\rm MH^-} , ~~(b) \hspace{0.25cm} {\rm M^{2-}}+{\rm Na^+} \xrightleftharpoons[\text{k$_{-n}$}]{\text{k$_n$}} {\rm MNa^-}, ~~(c) \hspace{0.25cm} {\rm MH^{-}}+{\rm H^+} \xrightleftharpoons[\text{k$_{-h2}$}]{\text{k$_{h2}$}} {\rm MH_2}, \nonumber \\
&&(d) \hspace{0.25cm} {\rm MNa^{-}}+{\rm Na^+} \xrightleftharpoons[\text{k$_{-n_2}$}]{\text{k$_{n_2}$}} {\rm MNa_2}, ~~(e) \hspace{0.25cm} {\rm MH^-} +{\rm Na^+} \xrightleftharpoons[\text{k$_{-hn}$}]{\text{k$_{hn}$}} {\rm MHNa}. \label{eqn:chem}
\een
We assume that all the binding sites/charge sites are identical and the binding affinities for the different ions are different. The species M$^{2-}$, MH$^-$, MNa$^-$, MH$_2$, MNa$_2$ and MHNa are different monomer species, all of which move with the polymer velocity.  The ion species are freely diffusible, but because they are ions, their movement is restricted by the requirement to maintain electroneutrality. Finally, because a small amount of water dissociates into hydrogen and hydronium, we are guaranteed that there are always some positive and negative ions in the solvent.

\subsection{Equations of motion and interface conditions} \label{EoM}
Suppose we have some volume $V$ of   a mixture comprised of $k$ types of particles
 each with particle density (number of particles per unit volume) $n_j$, and particle volumes $\nu_j$, $j = 1,\cdots, k$. From now on we will denote the quantities with subscript ``1'' related to PEG monomer species, subscript ``2'' related to ChS monomer species and subscript ``3'' related to solvent molecule. For each of these components there is a velocity, ${\bf v}_j$, the polymer network velocity ${\bf v}_1={\bf v}_2={\bf v}_p$ (the subscript ``p'' denotes polymer), the solvent velocity  ${\bf v}_3={\bf v}_s$ and molecular  species velocities ${\bf v}_j$, $j = 4,\cdots, k$. The volume fraction for each of the polymer species ($\theta_1, \theta_2$) and the solvent ($\theta_3 = \theta_s$) are $\theta_i=\nu_i n_i$, $i = 1, 2, 3$; respectively. The particle conservation of polymer species implies that
\ben
\frac{\partial \theta_i}{\partial t} + \nabla \cdot ({\bf v}_p \theta_i) = 0, \qquad i=1,2 \label{eq:CoM}
\een
Suppose the ChS monomers constitute a fraction `$\alpha$' of the total number of monomers, $n_m$,  (i.e., $n_2 = \alpha n_m$, $n_1 = (1-\alpha) n_m$) and the volume ratio of the ChS to PEG monomers is $\beta$ (i.e., $\nu_2 = \beta \nu_1$). Assuming that the other molecular species do not contribute significantly to the volume (i.e., we take $\nu_j=0$, $j=4,\cdots, k$), conservation of total volume implies $\theta_1+\theta_2+\theta_3 = 1$. From Eq. (\ref{eq:CoM}), it follows that
\ben
\nabla \cdot (\theta_s {\bf v}_s + (\theta_1+\theta_2){\bf v}_p) = \nabla \cdot \Big({\bf v}_s + \theta_1\Big\{\frac{1+(\beta-1)\alpha}{1 - \alpha}\Big\}({\bf v}_p-{\bf v}_s)\Big) = 0. \label{eq:1a}
\een
Further, in subsequent calculations we make an assumption that the density of ions particles is insignificantly small compared to the particle density of polymer and solvent, i.e. $\sum_{j\ge4} n_j \ll n_s+n_m$. The motion of the polymer and solvent phase of this multi-component mixture is governed by the Stokes equation for Newtonian fluid, which are
 \beq
\nabla\cdot( \theta_p \sigma_p({\bf v}_p)) - \xi \frac{\theta_s}{\nu_s} \phi_p ({\bf v}_p-{\bf v}_s) - \frac{\theta_p}{\nu_p} \nabla  \mu_p =0, \label{eq:motionVp}
 \eeq
and
\beq
 \nabla\cdot(  \theta_s \sigma_s({\bf v}_s)) -\xi\frac{{\theta_s}}{\nu_s}\phi_p  ({\bf v}_s-{\bf v}_p) - \frac{\theta_s}{\nu_s}\sum_{j\ge 4}  \hat{\phi}_j \nabla \mu_j -   \frac{\theta_s}{\nu_s}\nabla  {\mu_s} =0, \label{eq:motionVs}
  \eeq
where $\sigma_j({\bf v}) = \frac{\eta_j}{2} (\nabla  {\bf v}_j + \nabla  {\bf v}_j^T) + \lambda_j I\nabla\cdot {\bf v}_j$ is the   stress tensor ($\eta_j>0$ and $\lambda_j$,  $j=p,s$, are the  viscosities), $\xi_j$ are the drag coefficients, $\nu_p = (1-\alpha)\nu_1 + \alpha \nu_2$ is the averaged polymer particle volume, $\theta_p = \theta_1+\theta_2$, $\phi_s = \frac{n_s}{n_m+n_s}$, $\phi_p = \frac{n_m}{n_m+n_s}$, $\phi_j = \frac{n_j}{\sum_{i\ne 1,2}n_i}$ for $j\ge 4$ (assuming that the ions are dissolved in the solvent), are the polymer, solvent, and ion species per total solvent particle fractions, respectively. The third term in Eq. (\ref{eq:motionVs}) represents the force that the solvent feels from the ion chemical potentials. The ion species satisfy the force balance equation
\beq \xi_j  n_j ({\bf v}_s-{\bf v}_j)-n_j\nabla \mu_j=0, \qquad j\ge 4.\label{eq:ion_mom}
\eeq
In addition, if there is a free moving-edge to the gel, on one side of which (inside the gel) $\theta_p = \theta_p^-$, and on the other side of which (outside the gel)  $\theta_p^+ = 0$, $\theta_s^+ = 1$,  there are  interface conditions, 
 \beq
\sigma_p({\bf v}_p^-)  {\bf n}=  \frac{1}{\nu_p} \mu_p^-  {\bf n}  ,\label{eq:edge_Conditions_3}
 \eeq
for the polymer,  
\beq
 \left( \sigma_s({\bf v}_s^+)  -  \sigma_s({\bf v}_s^-)\right) {\bf n}   =\frac{1}{\nu_s}(\mu_s^+         - \mu_s^-) {\bf n}, \label{eq:edge_Conditions_4}
 \eeq
 for the solvent, and
 \beq
 \mu_j^+=\mu_j^-,\label{eq:ion_bc}
 \eeq for the  ion species, $j\ge 4$. {\bf n} is the normal to the free surface.
 The equations of motion and the interface conditions are derived using the standard variational arguments to minimize the rate of work, which constitutes the viscous rate of energy dissipated within the polymer and the solvent, energy dissipation rate due to the drag between solvent and polymer and between solvent and ion species particles as well as the rate of work required against the chemical potential, $\mu_j$. Readers are directed to \cite{Keener2011SIAM, Keener2011PRE, Keener2013} to see the derivation details.

\subsection{Chemical potential} \label{CP}
The chemical potentials, $\mu_j$, in Eq. (\ref{eq:motionVp}, \ref{eq:motionVs}) are calculated via the Gibb's free energy
\beq 
G = -k_BTS + U + PV,\label{eq:gibbs_fe} 
\eeq 
where $U$ is internal energy, $S$ is entropy, $T$ is temperature, $k_B$ is Boltzmann's constant, $P$ is pressure, and $V=\sum_j\nu_j n_j = \nu_1 n_1 + \nu_2 n_2 + \nu_s n_s$ is the total volume of the system (under the assumption that the volume occupied by the ions are small compared to the monomer and solvent volumes). The chemical potential are
\beq
\mu_j = \frac{\partial G}{\partial n_j} = -k_BT\frac{\partial S}{\partial n_j} + \frac{\partial U}{\partial n_j} + \nu_j P = \mu^S_j + \mu^I_j + \nu_j P, \label{eq:chem_pot}
\eeq
where $\mu^S_j$, $\mu^I_j$ are the contribution due to entropy and internal energy respectively.

\subsubsection{Entropic Contributions to Chemical Potentials} \label{CPEntropy}
The entropy of the system is defined as 
\beq 
S = \sum N_i\omega_i,
\eeq
where $\omega_i$ is the entropy per particle for the $i^{th}$ particle. Using standard counting arguments \cite{Doi1996}, for single-molecule species,
\beq \omega_j = -\ln(\phi_j) \qquad j \ge 3 
\eeq
%
%
%
%
The PEG and ChS chains exhibit permanent cross-linking bonds (i.e. covalent bonds), and the per-particle entropy is given by the rubber elasticity theory \cite{Doi2009}
\beq \omega_i = -\frac{3k_i}{2} \Big[ (\phi_i)^{-2/3} - 1 + \ln \phi_i \Big], \qquad i=1,2
\eeq
The particle fractions $\phi_1 = (1-\alpha) \phi_p$ and $\phi_2 = \alpha \phi_p$, $k_1, k_2$ are the fraction representing the number of cross-linked monomers in one chain of PEG and ChS, respectively. The entropic part of the chemical potentials are
%
%
%
\begin{align}
\frac{\mu^S_p}{{\it k}_B T} &=  k_1 \phi_1^{-2/3}\Big[\phi_1 + \frac{1-\alpha}{2}\Big] + k_2 \phi_2^{-2/3} \Big[\phi_2 + \frac{\alpha}{2}\Big] - \frac{3}{2}\Big[ k_2 \alpha (1+\phi_2\phi_s+\phi_2\phi_s\ln \phi_2) + \nonumber \\
&\hspace{0.5cm} k_1 (1-\alpha)(1+\phi_2\phi_s+\phi_2\phi_s\ln \phi_1) \Big] - \phi_s
\nonumber \\
\frac{\mu^S_s}{{\it k}_B T} &= \ln \phi_s + \phi_p - k_1 \phi_1 (\phi_1^{-2/3}+1) - k_2 \phi_2 (\phi_2^{-2/3}+1)  - \sigma_I \nonumber \\
\frac{\mu^S_j}{{\it k}_B T} &= \ln \phi_j + 1 - \sigma_I,
\end{align}
where $\sigma_I = \sum_{i\ge 4}\phi_j$ is the total ion particle fraction represents osmotic pressure as characterized by van't Hoff's law.

\subsubsection{Internal Energy Contribution to Chemical Potentials} \label{CPIntEnergy}
The internal energy consists of two contributions, long range electrostatic interactions and short range (nearest neighbor) interactions. The long range electrostatic interactions have energy
\beq
U_e = \sum_j z_jN_j\Phi_e,
\eeq
where $z_i$ is the charge on the $i^{th}$ ionic species ($z_1 \equiv z_p$ is the average charge per monomer), and $\Phi_e$ is the electric potential.

\begin{figure}[htbp]
\centering
\includegraphics[scale=0.18]{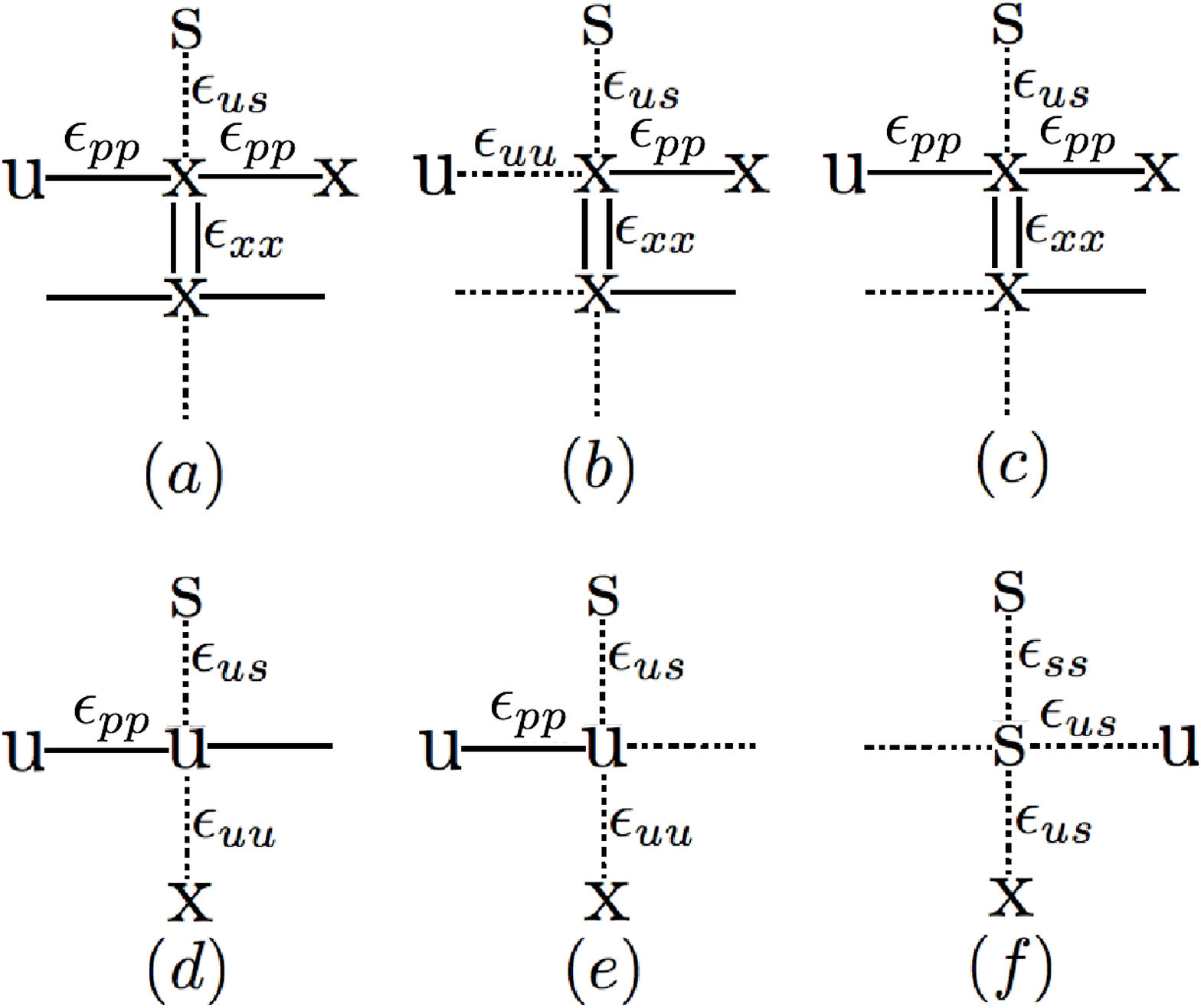}
\caption{Pairwise interactions of the different species and their associated interaction energies for: (a)-(c) cross-linked monomers, x,  (d) and (e) uncross-linked monomers, u, and (f) solvent, s. Double solid lines denote cross-linking, single solid lines denote interactions within a polymer chain while the other types of interactions are denoted by dashed lines. The nearest neighbor interactions are assumed identical for both the PEG and ChS monomer species.} \label{fig:Fint}
\end{figure}

To calculate the short range interaction energy for the polymer and solvent, we assume that for each of the $n_T$ (= $n_m + n_s$) particles have $z$ neighboring interaction sites (called the coordination number). Of the total of $n_m$ monomers, $k_1 n_1 = k_1 (1-\alpha) n_m$ and $k_2 n_2 = k_2 \alpha n_m$ of them are the cross-linked PEG and ChS pairs, respectively, and since the cross-linked particles are connected, we treat them as single species. The different species with their pairwise interactions (for either of the polymer species) are shown in Fig.~\ref{fig:Fint}. The cross-linked particle pairs have 2$z$-6, 2$z$-4 or 2$z$-5 free interaction sites, depending on whether the monomers in the cross-linked pair are   both in the middle of a polymer chain,   both at the end of a chain, or have one monomer in the middle and the other at the end of a polymer chain, respectively (Fig.~\ref{fig:Fint}a,b,c). An uncross-linked monomer has either $z$-2 or $z$-1 free interaction sites, based on its position in the polymer chain (Fig.~\ref{fig:Fint}d,e), while the solvent particles have $z$ free interaction sites (Fig.~\ref{fig:Fint}f). 

We let $k_B T_0 \epsilon_{xx}$, $k_B T_0 \epsilon_{uu}$, $k_BT_0\epsilon_{pp}$, $k_B T_0 \epsilon_{us}$ and $k_B T_0 \epsilon_{ss}$ be the interaction energy associated with the covalent cross-links, uncross-linked monomer-monomer interaction (with the assumption that the different types of monomers have the same interaction energy), monomer-monomer interaction within a polymer strand, monomer-solvent interaction and solvent-solvent interaction, respectively. These interactions are assumed identically equal for both PEG and ChS monomers. $T_0$ is a reference temperature. The interaction energies for the different species of polymer and solvent particles, calculated using standard mean-field arguments, are 
\begin{eqnarray}
\frac{F^1_x}{k_B T_0} &=& r \left(\frac{k_1N_1-2}{k_1N_1}\right)\left(\frac{k_2N_2-2}{k_2N_2}\right) n_m \left(2\epsilon_{xx}+(2z-6)\Eus+ 4\epsilon_{pp}\right), \\
\frac{F^2_x}{k_B T_0} &=& r \left(\frac{2}{k_1N_1}\right)\left(\frac{2}{k_2N_2}\right) n_m \left(2\epsilon_{xx}+(2z-4)\Eus + 2\epsilon_{pp} \right), \\
\frac{F^3_x}{k_B T_0} &=& r \left(\frac{2}{k_1N_1}\frac{k_2N_2-2}{k_2N_2} + \frac{2}{k_2N_2}\frac{k_1N_1-2}{k_1N_1}\right) n_m \left(2\epsilon_{xx}+(2z-5)\Eus +3\epsilon_{pp}\right), \\
\frac{F^1_u}{k_B T_0} &=& \left((1-k_1)(1-\alpha)\frac{k_1N_1-2}{k_1N_1} + (1-k_2)\alpha\frac{k_2N_2-2}{k_2N_2}\right)n_m \left( (z-2)\Eus + 2\epsilon_{pp}\right), \\
\frac{F^2_u}{k_B T_0} &=& \left((1-k_1)(1-\alpha)\frac{2}{k_1N_1} + (1-k_2)\alpha\frac{2}{k_2N_2}\right)n_m \left( (z-1)\Eus + \epsilon_{pp}\right), \\
\frac{F_s}{k_B T_0} &=& zn_s \left(\epsilon_{us} \frac{N_m}{N_T} + \epsilon_{ss}\frac{N_s}{N_T}\right),
\end{eqnarray}
respectively, where $r = \frac{k_1(1-\alpha)+k_2\alpha}{2}$ and $\Eus =  \epsilon_{uu} \frac{n_m}{n_T} + \epsilon_{us}\frac{n_s}{n_T} $. The total per particle interaction energy is
\begin{align}
U^I &= \frac{1}{2(n_m+n_s)}(F^1_x+F^2_x+F^3_x+F^1_u+F^2_u+F_s)\nonumber
 \\
 &= k_B T_0 \left[\chi \phi_p\phi_s + \mu_0^s \phi_s+\mu_0^p\phi_p + \frac{z}{2}\epsilon_{us} \right],\label{eq:29}
\end{align}
where
\begin{align}
\chi &= \frac{z}{2}(\epsilon_1 + \epsilon_2) - \Big( r + 1+ \frac{k_1(1-\alpha)-k_2\alpha-2(1-\alpha)}{2k_1N_1} + \frac{k_2\alpha-k_1(1-\alpha)-2\alpha}{2k_2N_2} \Big)\epsilon_1, \nonumber \\ 
\mu_0^p &= r \epsilon_3 -\frac{z}{2}\epsilon_1 + \Big( 1+ \frac{k_1(1-\alpha)-k_2\alpha-2(1-\alpha)}{2k_1N_1} + \frac{k_2\alpha-k_1(1-\alpha)-2\alpha}{2k_2N_2} \Big)\epsilon_4, \nonumber \\
\mu_0^s &= -\epsilon_{2}\frac{ z }{ 2 }, \label{eqn:mu0}
\end{align}
and 
\beq
\epsilon_1 = \epsilon_{us}-\epsilon_{uu},~~\epsilon_2 = \epsilon_{us}-\epsilon_{ss},~~\epsilon_3 = \epsilon_{xx}-\epsilon_{uu},~~\epsilon_4 = \epsilon_{pp}-\epsilon_{uu}. \label{eq:epsilons}
\eeq
In polymer chemistry literature, the coefficients, $\chi, \mu_0^p, \mu_0^s$, are referred as the Flory interaction parameter and the chemical potentials of pure polymer and solvent species respectively \cite{Flory1953}. The factor $\frac{1}{2}$ in the per-particle interaction energy (Eqn. \ref{eq:29}) is to correct for double counting. The corresponding contributions to chemical potentials are 
 \beq
 \mu_p^I=k_B T_0 \left( \chi\phi_s^2   +\mu_p^0 \right), \qquad  \mu_s^I=k_B T_0 \left(\chi \phi_p^2  + \mu_s^0 \right).
 \eeq
In summary, the chemical potential for the polymer and solvent phase is
\ben
\frac{\mu_j}{{\it k}_B T} = M_i + z_i \Psi_e + \nu_i \frac{P}{{\it k}_B T}, \qquad j = p,s,
\een
where $\Psi_e = \frac{\Phi_e}{k_BT}$, $M_p = \frac{1}{{\it k}_B T} (\mu^S_p + \mu^I_p)$ and $M_s = \frac{1}{{\it k}_B T} (\mu^S_s + \mu^I_s) + \sigma_I$. The functions $M_i$ are chosen so as to highlight the contribution of the various terms in the net swelling pressure (see Eq. (\ref{eq:net_swelling}) later). For the ion species,  $\mu_j^I=0$, and we ignore its volume, $\nu_j$, so that  
\beq \frac{\mu_j}{k_BT} =  \ln\phi_j+1-\sigma_I + z_j\Psi_e, \hspace{1cm} j\ge 3,\label{eq:ion_Chem_Pot}
\eeq
Finally, the gel-sol interface conditions, Eqns.~(\ref{eq:edge_Conditions_3}-\ref{eq:edge_Conditions_4}), become 
\begin{align}
 \sigma_p({\bf v}_p^-)  {\bf n} &= \frac{k_BT}{\nu_p} \Big( M_p^-+   z_p\Psi_e + \nu_p\frac{P}{k_BT} \Big){\bf n}, \nonumber \\
 \left(\sigma_s({\bf v}_s^+) - \sigma_s({\bf v}_s^-)\right){\bf n} &= \frac{k_BT} {\nu_s}\Big(M_s^+ - M_s^- -\sigma_I^+         +\sigma_I^- -\nu_s\frac{P}{k_BT}\Big) {\bf n} ,\label{eq:edge_Conditions_6}
 \end{align}
where $P\equiv P^-$, $P^+=0$ and $\Psi_e^-\equiv \Psi_e$, $\Psi_e^+=0$.  Eliminating $P$ from these we find the single interface condition
   \beq
 \left(\sigma_p({\bf v}_p^-)-\sigma_s({\bf v}_s^-)+  \sigma_s({\bf v}_s^+)\right){\bf n}= \Sigma_{net}{\bf n}, \label{eq:ic}
 \eeq
 where $\Sigma_{net}$ is the net swelling pressure,
 \beq
 \frac{\Sigma_{net}}{k_BT}=  \frac{M_p^-}{\nu_p}-\frac{M_s^-}{\nu_s}+ \frac{T_0}{T}\frac{\mu_s^0}{\nu_s}+  \frac{z_p}{\nu_p}\Psi_e +\frac{\sigma_I^-}{\nu_s} -\frac{\sigma_I^+}{\nu_s } .        \label{eq:net_swelling}
 \eeq
 The term $(z_p \Psi_e)$ represents the Donnan swelling pressure while $(\sigma^-_I - \sigma^+_I)$ represents the Osmotic pressure due to the ions dissolved in the solvent (also termed as the `Net Osmolarity' inside the gel).
 
\subsection{Ionized-species chemistry} \label{Chem}
 
Let the concentrations per total volume of the polymer species be denoted by $m=[\text{M}^{2-}], y=[\text{MH}^-], v=[\text{MNa}^-], w=[\text{MH}_2], x=[\text{MNa}_2], q=[\text{MHNa}]$, with the total monomer concentration 
 \beq m_T = m + y + v + w + x + q, \label{eq:47}
 \eeq
and the ChS-monomer volume fraction $\theta_2 = \nu_2 N_A m_T$. The charged ChS monomer concentration, m, is obtained from Eqn. (\ref{eq:47}). The concentrations per solvent volume of the ion species are denoted as $n = [{\rm Na^+}]$, $h = [{\rm H^+}]$,  and $c_l = [{\rm Cl^-}]$. With concentrations expressed in units of moles per liter, the relationship between ion particle fractions $\phi_j$ and concentrations $c_j$ is $\phi_j = \nu_sN_A c_j$, where $N_A$ is Avagadro's number. Under the assumption of fast chemistry, the law of mass action for the monomer binding reactions (Eq.(\ref{eqn:chem})) reduces to
\begin{align}
& (a) \qquad k_h m h \theta_s + ( k_{-h_2} w + k_{-hn} q ) \phi^2_s = k_{-h} y \phi^2_s + (k_{h_2} y h  + k_{hn} y n) \theta_s \nonumber \\
& (b) \qquad (k_n m n + k_{n_2} v n) \theta_s = (k_{-n} v + k_{-n_2} x) \phi^2_s \nonumber \\
& (c) \qquad k_{h_2} y h \theta_s = k_{-h_2} w \phi^2_s \nonumber \\
& (d) \qquad k_{n_2} v n \theta_s = k_{-n_2} x \phi^2_s \nonumber \\
& (e) \qquad k_{hn} y n \theta_s = k_{-hn} q \phi^2_s \label{eqn:massAction}
\end{align} 
Since the unbinding (dissociation) reactions are ionization reactions that require two ``units" of solvent, we take the unbinding reaction rates to be $ k_{-C}\phi_s^2$, $C = h, n, h_2, n_2, hn$, and because ChS monomers carry a double negative charge, $2k_{i_2}=k_{i}, k_{-i_2} = 2k_{-i}$, $i = n, h$ and $4k_{hn}=k_h+k_n, k_{-hn} = k_{-h}+k_{-n}$. Simplifying Eq. (\ref{eqn:massAction}),
\beq
y = \frac{\theta_s}{ K_h\phi_s^2}  mh, ~~ v =  \frac{\theta_s}{ K_n\phi_s^2}  mn, ~~ w = \Big(\frac{\theta_s}{ 2 K_h\phi_s^2}\Big)^2  mh^2, ~~ x = \Big(\frac{\theta_s}{ 2 K_n\phi_s^2}\Big)^2  mn^2, 
q = \Big(\frac{\theta_s}{ \sqrt{K_h K_{hn}}\phi_s^2}\Big)^2  mhn,
\eeq
where $K_h = \frac{k_{-h}}{k_h}$, $K_n = \frac{k_{-n}}{k_n}$, $K_{h_2} = 4 K_h$, $K_{n_2} = 4 K_n$ and $K_{hn} = 4 \frac{k_{-h} + k_{-n}}{k_h + k_n}$.

Similarly, when the diffusion of the ion species and their binding and unbinding reactions are fast compared to the swelling kinetics of the polymer network, the Nernst-Planck equation, describing the ion movement (together with the interface conditions Eq.(\ref{eq:ion_bc})), reduces to
\beq C = C_b e^{-z_C \Psi_e} \label{eq:ion_bal3c} \eeq
with $C = h, n, c_l$, $z_n= z_h=1$ and $z_{c_l} = -1$ and the subscript `b' denotes the corresponding bath concentrations. The electrostatic potential, $\Psi_e$, is determined by the electroneutrality constraint inside the gel,
 \beq  (n + h-c_l)\theta_s + z_p m_T=0, \label{eq:el_bal} \eeq
where $z_p$ is the average charge per ChS-monomer. The electro-neutrality in the bath requires (i.e. setting $\Psi_e=0$ in Eqn.~\ref{eq:ion_bal3c} and then using Eqn.~\ref{eq:el_bal})
\beq 
n_b + h_b - {c_l}_b = 0, \label{eq:bath}.
\eeq
The average charge, $z_p$, depends on the residual charge of the bound ChS-polymer species,
\beq z_p m_T = -(2 m + y + v). \eeq
The system of equations including the mass conservation Eqn.~(\ref{eq:CoM}), total volume conservation Eqn.~(\ref{eq:1a}) together with force balance Eqns.~(\ref{eq:motionVp}-\ref{eq:motionVs}) and interface condition Eqn.~(\ref{eq:ic}), subject to the constraints  Eqn.~(\ref{eq:47})  (monomer conservation), Eqn.~(\ref{eq:ion_bal3c}) (ion motion) and Eqn.~(\ref{eq:el_bal}) (electroneutrality); completely describes the dynamical motion of a freely swelling gel. Finally, the `Net-Osmolarity' in the gel, Osm, is 
\beq {\rm Osm} = (n + h + c_l) - (n_b + h_b + c_{l_b}) = (n_b + h_b)(e^{-\Psi_e} + e^{\Psi_e} - 2),\eeq
which measure the excess moles of solute (inside the gel) per liter of solvent.

\section{Material and methods }\label{sec:MM}
Poly(ethylene glycol) dimethacrylate (PEGDM) was synthesized via microwave methacrylation \cite{Lin-Gibson2004} by reacting poly(ethylene glycol) (Mol. wt. 4600 g/mol) with methacrylic anhydride in the presence of hydroquinone (Sigma-Aldrich). The PEGDM product was recovered by dissolving the mixture in methylene chloride followed by precipitation with ethyl ether, filtration and drying. The degree of methacrylation was determined to be 85\% by $^1$HNMR (Varian VYR-500). Methacrylated chondroitin sulfate (ChSMA) was synthesized as described previously \cite{Bryant2004} by reacting chondroitin sulfate A (Sigma-Aldrich) in a 1:8 ratio with methacrylic anhydride in deionized water (DI-H2O) at 4$^o$C and a pH of 8 for 24 hours. The product was recovered by precipitation in chilled methanol and dialyzed in DI-H$_2$O overnight. The purified product was recovered via lyophilization and the degree of methacrylation was determined to be 16\% by $^1$HNMR (Varian VYR-500), indicating that there were 16 methacrylates per ChSMA molecule. 

Ionic gels were formed by co-polymerizing PEGDM macromers with ChSMA macromers at a final macromer concentration of 10 \% (g/g) with 0.05\% (g/g) photoinitiator Irgacure 2959 (Ciba Specialty Chemicals) in deionized water (DI-H$_2$O).  The macromer solution was placed in between two glass slides with a 1 mm spacer and exposed to 365 nm light with an intensity of $~$5 mW cm$^2$ for ten minutes.  The ratio of PEGDM:ChSMA in the macromer solution was varied as follows: 90:10, 80:20, 70:30, 60:40 and 50:50 by weight. After polymerization, cylindrical gels were punched out using a 5 mm diameter biopsy punch and immediately weighed before being put into a solution and allowed to swell.  The gels were then placed in the solution of interest to investigate the effect of salt concentration and pH on the swelling of the PEGDM:ChSMA gels.  In a separate experiment, a subset of gels was placed in DI-H$_2$O for 48 hours and the soluble fraction of the ChSMA (i.e., the fraction not incorporated into the gel) was assessed using the DMMB dye assay \cite{Barbosa2003}. It was confirmed that 99\% of the ChSMA was incorporated into the gel. 

Gel swelling in electrolyte solutions of varying concentrations of NaCl in DI-H$_2$O was investigated. Solutions were prepared of 0.1 mM, 1 mM, 10 mM, 0.1M, and 1 M NaCl while the pH was maintained at pH 7. The gels were placed into each solution after the initial mass was taken and allowed to swell for 48 hours. Gel mass was measured at various time points throughout the 48 hours to ensure that gels had reached their equilibrium swelling by 48 hours.  The effect of pH on the swelling of PEGDM:ChSMA gels was also explored. These gels were placed in various pH solutions, which were prepared using 1M NaOH and 1M HCl in DI water. The pH of the solvents varied from pH 3 to pH 8. An initial mass of the gel was taken for a 5-mm disk immediately after polymerization. Gels were placed in the solution of interest and mass measured at various time points up to 48 hours to ensure equilibrium swelling was reached. 

The initial volume of the dry polymer, V$_i$, was determined from the initial mass, m$_i$, using a weighted density of the ChSMA, PEGDM, and DI-H$_2$O, as follows
\beq
V_i = 0.1 \Big( \frac{m_i \cdot \% PEGDM}{\rho_{PEGDM}} + \frac{m_i \cdot \% ChSMA}{\rho_{ChSMA}} \Big) 
\eeq
The mass of the gel at equilibrium was measured and used to determine the swollen volume V$_{swollen}$. It was assumed that all polymer was incorporated in the gel, which was confirmed for chondroitin sulfate, and therefore a change in mass was solely due to the a change in the water content. Hence
\beq
V_{swollen} = 0.1 \Big( \frac{m_i \cdot \% PEGDM}{\rho_{PEGDM}} + \frac{m_i \cdot \% ChSMA}{\rho_{ChSMA}} \Big) + \frac{m_{swollen}-10\% m_i}{\rho_{solvent}}
\eeq
The percent of PEGDM and ChSMA was dependent on the ratio of PEGDM:ChSMA which varied from 90:10 to 50:50. The initial mass, m$_i$, was the mass of the gel immediately following polymerization and before being placed in a solvent. The swollen mass, m$_{swollen}$, is the mass of the gel at equilibrium. The density of the PEGDM, $\rho_{PEGDM}$, was estimated to be 1.07 g/ml and the density of the ChSMA, $\rho_{ChSMA}$ , was assumed to be 1.001 g/ml \cite{Milch1965}. The density of the solvents, $\rho_{solvent}$, was assumed to be 1.00 g/ml. The volume fraction of the polymer was calculated as the volume of the dry polymer (i.e. before swelling) divided by the volume of the swollen gel, $V_i / V_{swollen}$.

\section{Results and discussion }\label{sec:results}
At equilibrium, we solve the system of equations given by the interface conditions (Eqn.~\ref{eq:ic}), monomer conservation (Eqn.~\ref{eq:47}), ion motion (Eqn.~\ref{eq:ion_bal3c}) and electroneutrality (Eqn.~\ref{eq:el_bal}). The mass, volume and the force balance (Eqns.~\ref{eq:CoM}-\ref{eq:motionVs}) are trivially satisfied. The values of the parameter used in our numerical calculations are listed in Table~\ref{tab:Table1}. The constants in the model are the monomer volumes, $\nu_1, \nu_2$, the number of nearest neighbor in the PEG-ChS polymer lattice, $z$, and the nearest neighbor interaction energies, $\epsilon_i$, assumed identically equal for both types of polymer (eqn.\ref{eq:epsilons}). The undetermined constants are the binding affinities of the various
cations with the gel, $K_h, K_n$ (eq.\ref{eqn:massAction}).

The monomer volumes are found from the density and the molecular weight information ($\nu_i = M_i/(\rho_i * N_A)$). The cross-linked PEG-ChS matrix has a 3-D configuration, which suggests that we choose the coordination number, $z=6$, hence mimicking a 3-D cubic lattice. The interaction energies are found from the solubility data, $\delta_i$, and the standard free energies, ${k}_B T_0 \mu^0_p$ and ${k}_B T_0 \mu^0_s$ (eqn.\ref{eq:29}). The standard free energy, is the energy of all the interactions between the molecule and its neighbors in a pure state that have to be disrupted to remove the molecule from the pure state \cite{Barton1990} (pg.141-144). These can be determined from the Hildebrand solubility parameters, $\delta_i$, via
\begin{align}
-{\it k}_B T_0 \mu^0_p (\alpha=0) &= \nu_1\delta^2_1 \nonumber \\
-{\it k}_B T_0 \mu^0_p (\alpha=1) &= \nu_2\delta^2_2 \nonumber \\
-{\it k}_B T_0 \mu^0_s &= \nu_w\delta^2_w, \label{eq:HSP}
\end{align}
where $\nu_w$ is the volume of 1 molecule of water. The negative sign in eqn.\ref{eq:HSP} indicates that ${\it k}_B T_0 \mu^0_p, {\it k}_B T_0 \mu^0_s < 0$, since they are the interaction energies. For example, at $T_0 = 298$K, $\nu_w = 2 \times 10^{-23}$ cm$^3$ as the monomeric volume of 1 water molecule and the solubility parameter $\delta_w=48.07 MPa^{1/2}$, the non-dimensional parameter, $\mu^0_s$ (from eq.\ref{eq:29}), is
\beq \mu^0_s = -\epsilon_2 \frac{z}{2} = -\nu_m \delta^2 / ({\it k}_B T_0) = -11.23,
\eeq
giving the value of the interaction energy, $\epsilon_2=3.74$.
\begin{table}[htbp]
\centering
\begin{tabular}{|c|c|c|c|c|}
\hline
 & PEG ($i=1$) & ChS ($i=2$) & Units & Source\\
\hline
Density ($\rho_i$) & 1.07 & 1.001 & g/mL & \cite{Bryant2004, Milch1965} \\
\hline
Molecular weight (M$_i$) & 4600 & 48700 & g/mol & \cite{Bryant2005}  \\
\hline
Repeat unit per chain ($N_i$) & 102 & 86 & -- & \cite{Bryant2005} \\
\hline
Cross-link fraction ($k_i$) & 0.25, 0.5, 0.75 & 0.25, 0.35, 0.45 & -- & \cite{Bryant2003} \\
\hline
Hildebrand solubility ($\delta_i$) & 17.39 & 5.19 & MPa$^{1/2}$ & \cite{Barton1990} \\
\hline
\multicolumn{2}{|l|}{Hildebrand solubility for water ($\delta_w$)} & 48.07 & MPa$^{1/2}$ & \cite{Barton1990} \\\hline
\end{tabular}
\caption{Parameters common to all the numerical results. The reference temperature for the solubility parameters is fixed at $T_0=298$K.}\label{tab:Table1}
\end{table}
%
%
%
%
%
%
\subsection{Parameter estimation} \label{subsec:PE}
\begin{figure}[htbp]
\centering
\subfigure[]{\includegraphics[scale=0.5]{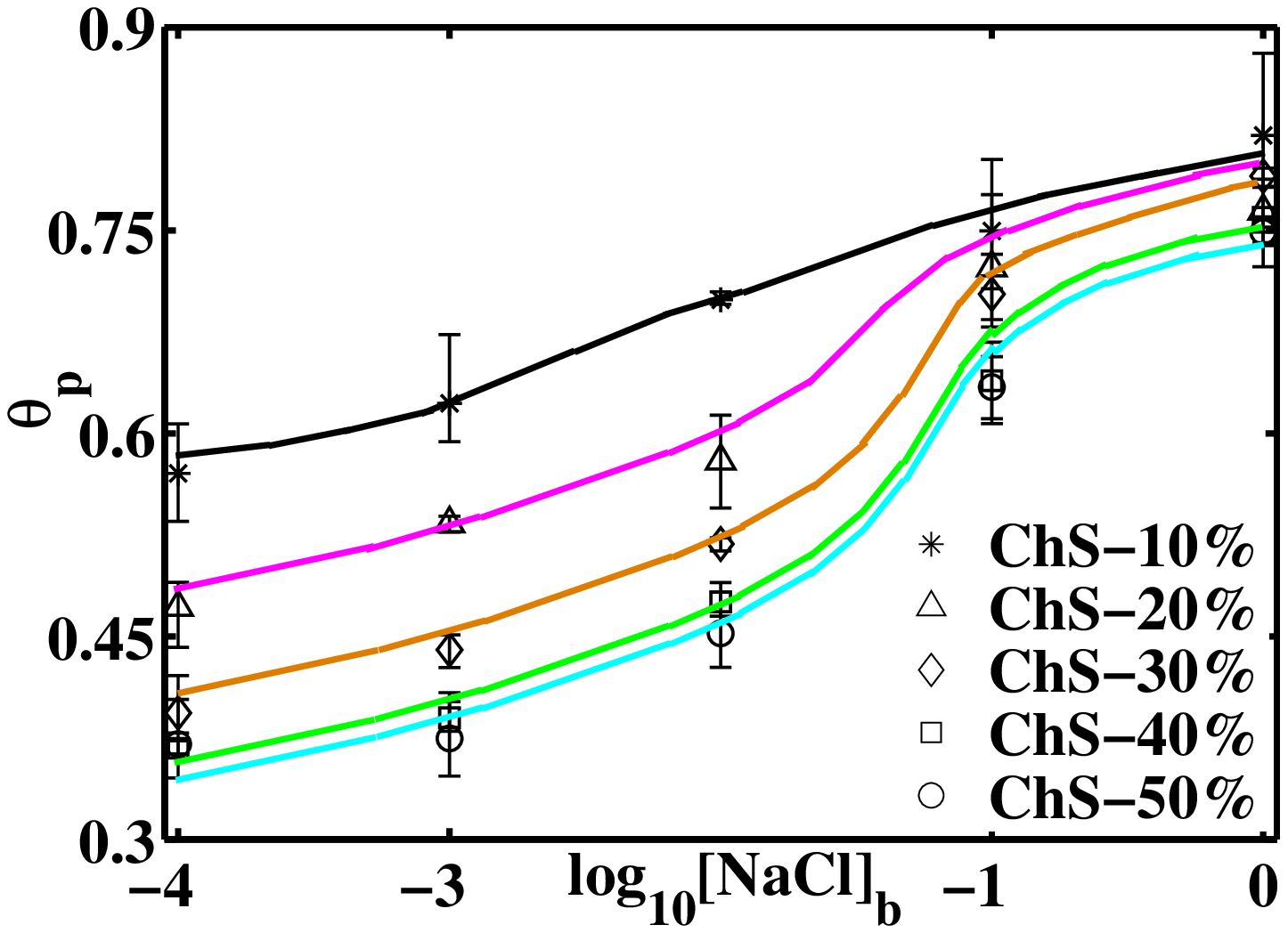}}
\subfigure[]{\includegraphics[scale=0.5]{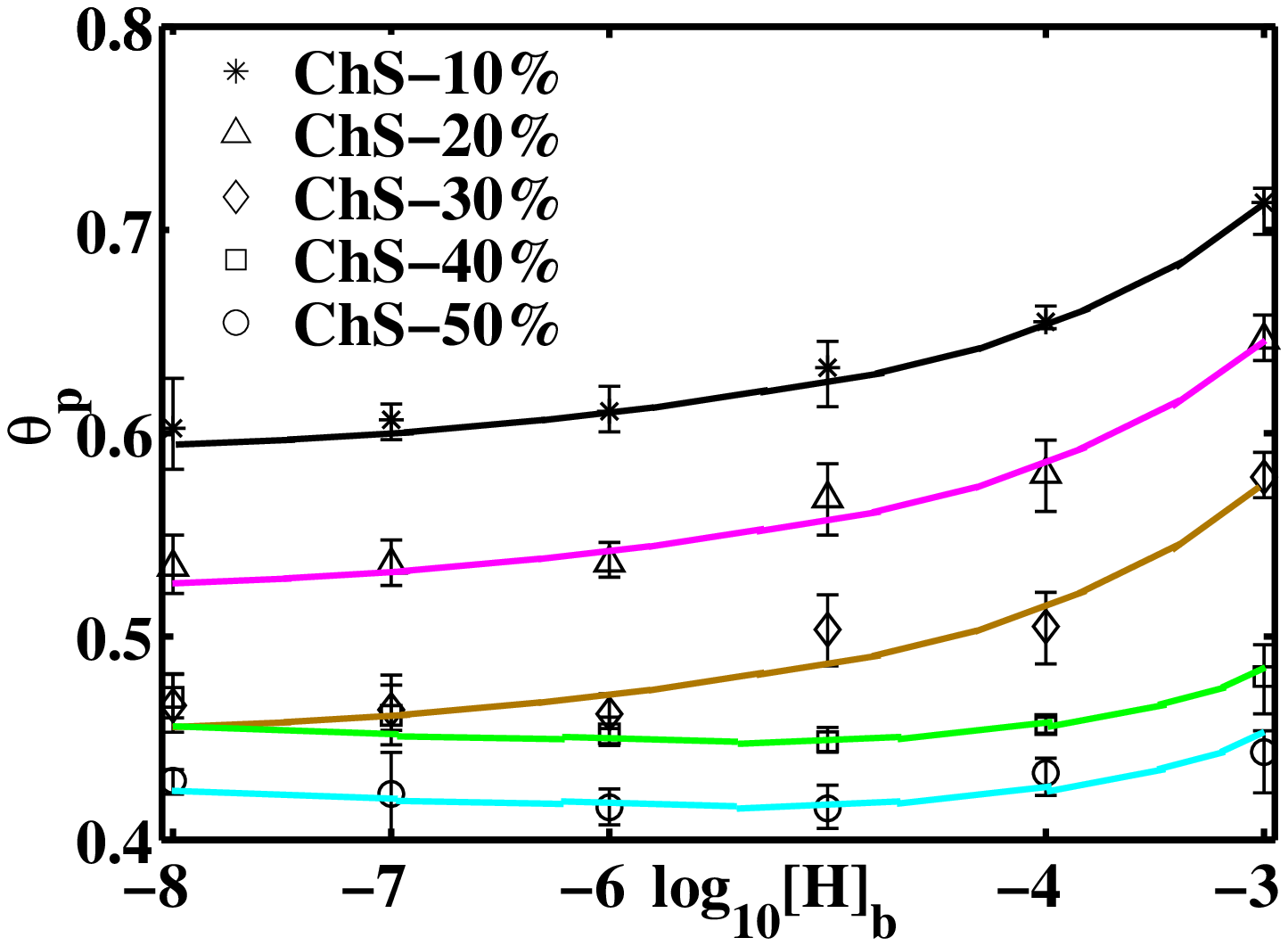}}
\caption{Experimental data of equilibrium volume-fraction of the PEG-ChS gel vs. (a) different NaCl concentration in the bath (in mol/lt) and pH = 7.0, and (b) different pH and zero salt concentration. The sample is a 10 wt\% gel. The molecular mass and the density of PEG 4600 g/mol and 1.07 g/mL, while that of ChS is 48700 g/mol and 1.001 g/mL, respectively. The experiments were performed at a constant room temperature of 293K. The solid lines correspond to the volume fraction values predicted by the model.}\label{fig:Fig2}
\end{figure}

The experimental data are used to calibrate the model for the gel volume-fraction at equilibrium \cite{Bryant2004}. Fig~\ref{fig:Fig2} presents the sample averaged equilibrium data-points at different salt concentrations and neutral pH (Fig~\ref{fig:Fig2}a) and at variable pH solution (Fig~\ref{fig:Fig2}b). The equilibrium values were noted at time $t=48$ hrs. Each point represents the average of four samples, with values noted in identical conditions, with the upper and the lower limits in the error bar representing maximum and the minimum variation from the average, respectively.

The molecular mass and the density of PEG were fixed at 4600 g/mol and 1.07 g/mL, while that of ChS were fixed at 48700 g/mol and 1.001 g/mL, respectively. These values give the monomer volumes of the PEG chains as $\nu_1=7.14 \times 10^{-21}$ cm$^3$, and that of ChS chains as $\nu_2=8.08 \times 10^{-20}$ cm$^3$. The Hildebrand solubility parameters for pure species (values given in Table~\ref{tab:Table1}) and the monomer volumes are used to calculate the interaction energies, $\epsilon_i$ (i=1,...,4). Using the relations in Eqn.~(\ref{eq:HSP}), these values are fixed at $\epsilon_1=0.0, \epsilon_2=3.74, \epsilon_3=9.58, \epsilon_4=-56.17$. The reference temperature is fixed at $T_0=298$K, while experiments were performed at $T=293$K. The undetermined parameters, namely the binding affinities, $K_h, K_n$ (Eqn.\ref{eqn:massAction}) are computed by constructing a nonlinear least-square function and employing the conventional subspace trust region method implemented in MatlabÕs {\bf lsqnonlin} \cite{Coleman1994}. These values are found as log$_{10}(K_n)=-2.51$, log$_{10}(K_h)=-3.56$. It is observed that the cross-link fraction of these gels is variable under identical experimental conditions. Therefore, we study the effect of cross-linking on the equilibrium configuration by selecting 3 different pairs of cross-link fractions for the PEG-ChS gels: CL$_1$: (0.25, 0.25), CL$_2$: (0.5, 0.35) and CL$_3$: (0.75, 0.45). The solid lines (in Fig.~\ref{fig:Fig2}) are the equilibrium values predicted by the model.
%
%
%
%
\subsection{Effects of changes in the gel composition} \label{subsec:GC}
\begin{figure}[htbp]
\centering
\subfigure[]{\includegraphics[scale=0.5]{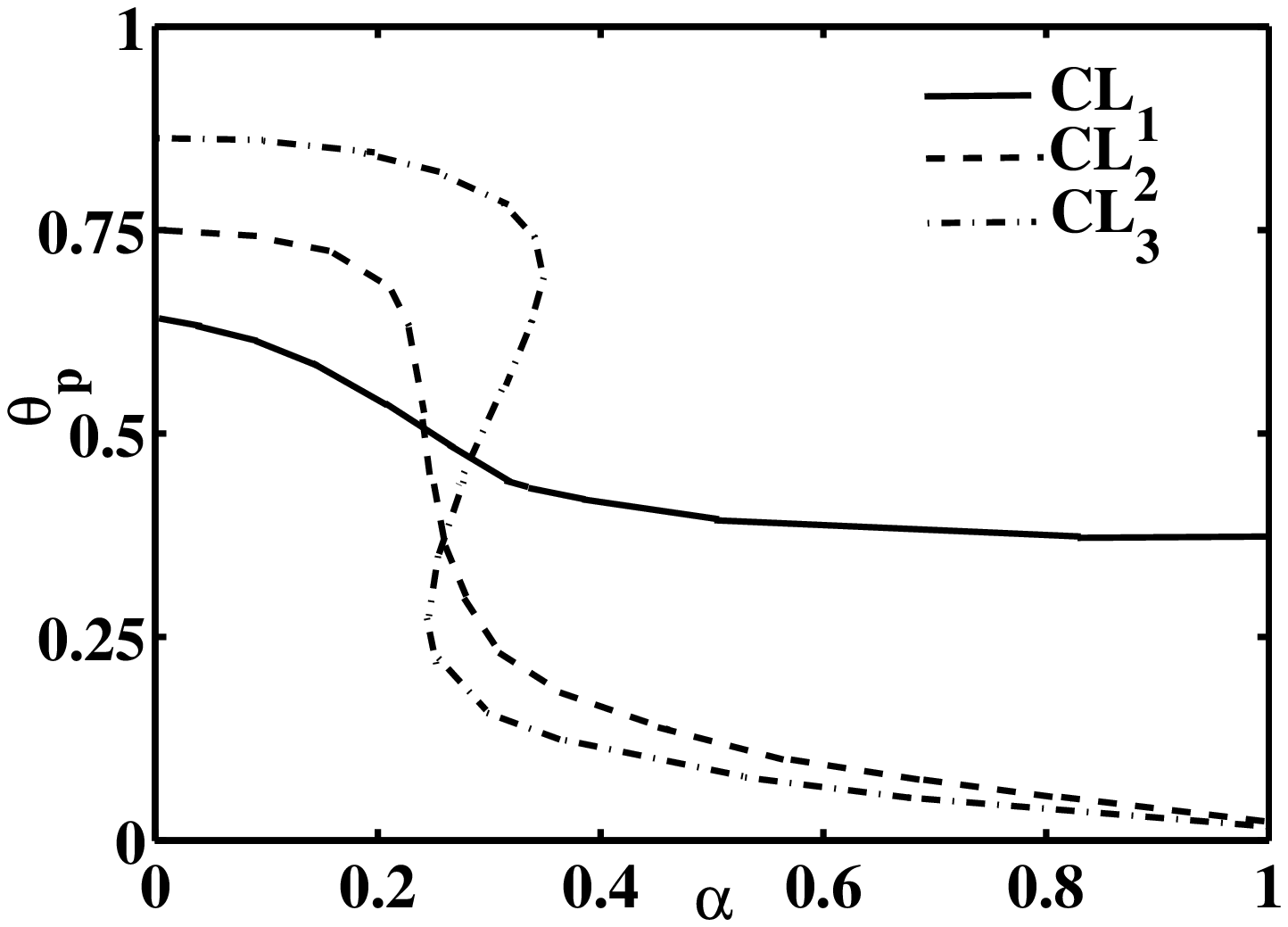}}
\subfigure[]{\includegraphics[scale=0.5]{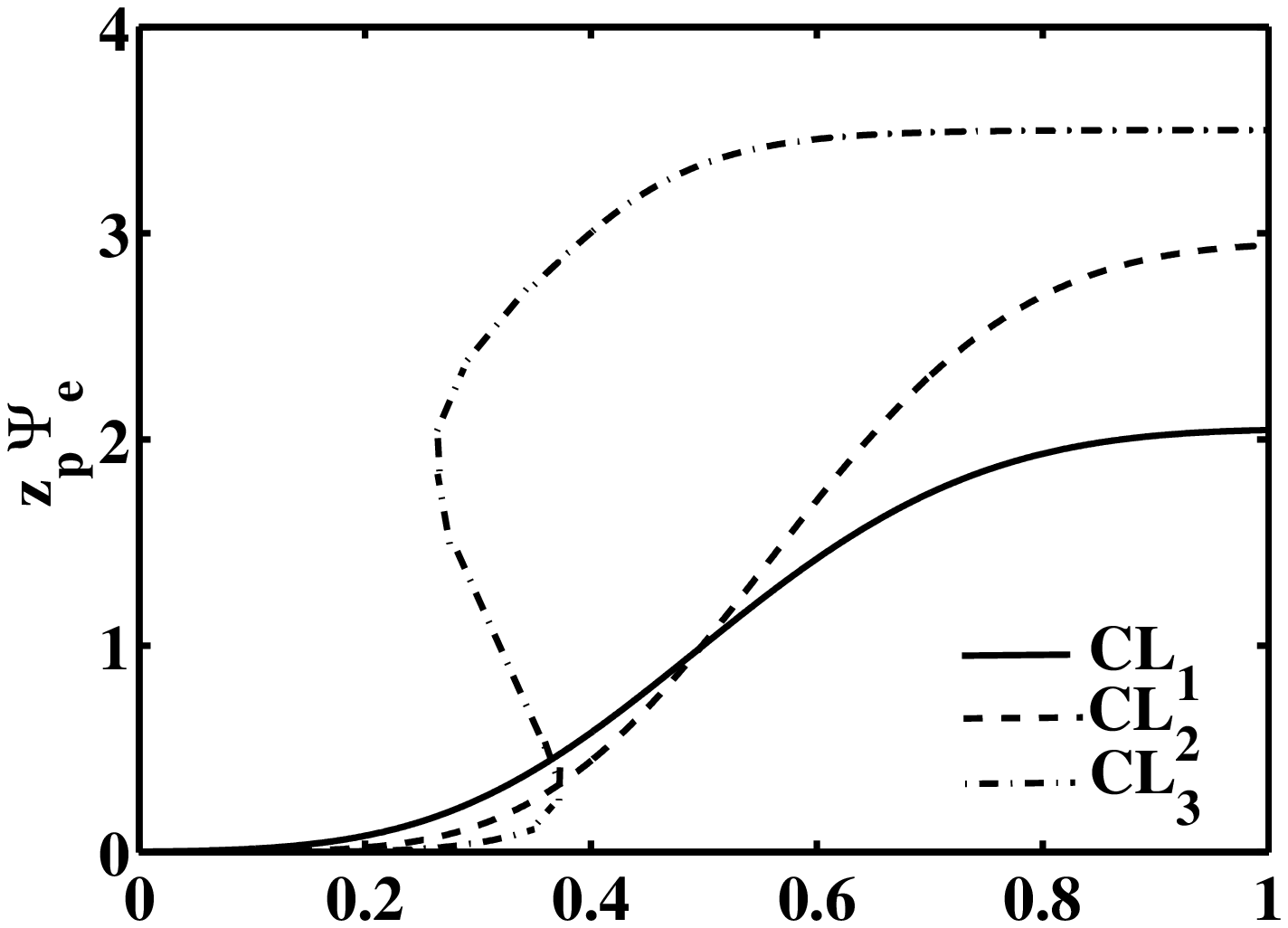}}
\vskip 0.0001cm
\subfigure[]{\includegraphics[scale=0.5]{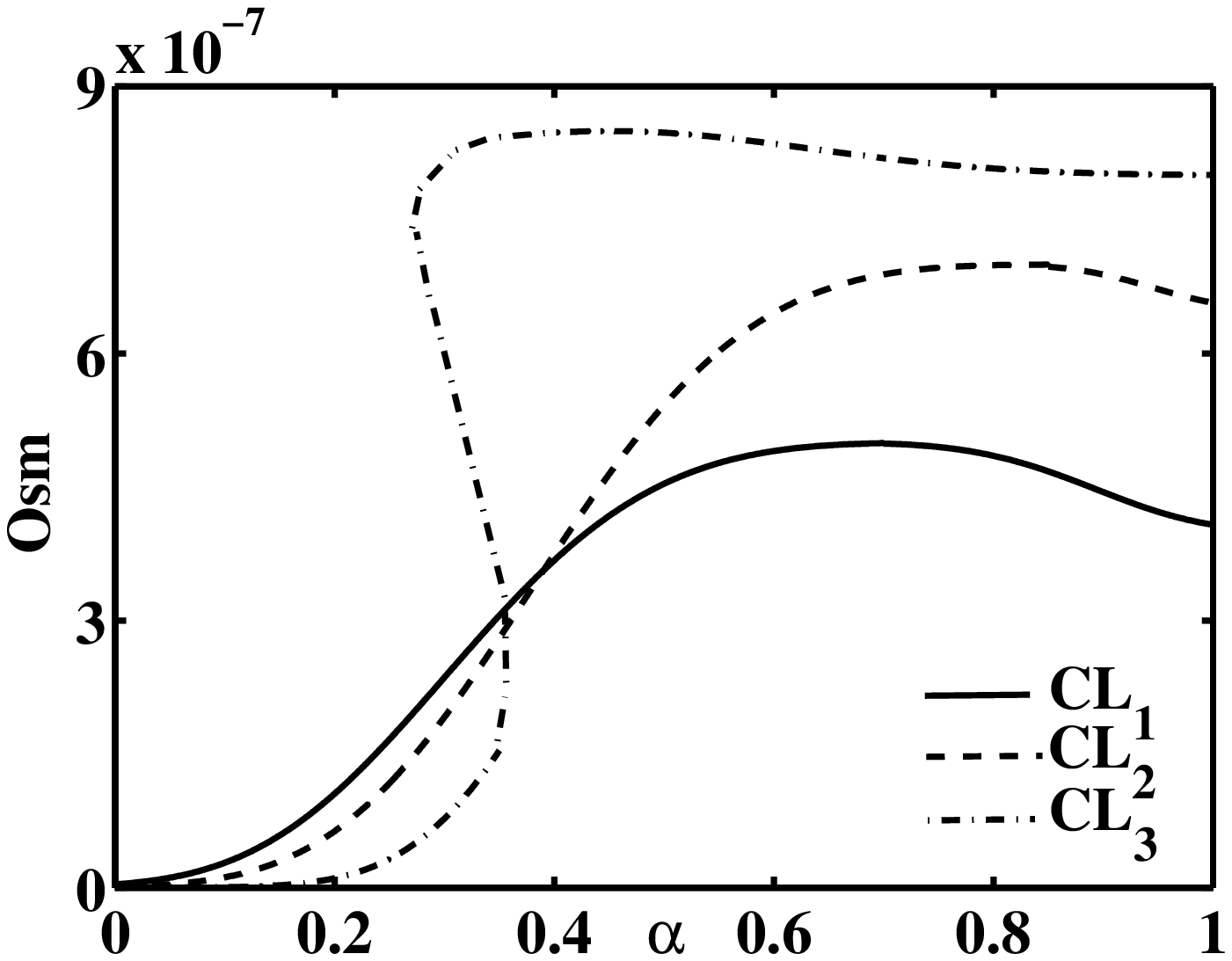}}
\caption{(a) Equilibrium polymer volume-fraction, (b) Donnan swelling pressure, and (c) Net-Osmolarity, vs. $\alpha$, the fraction of Chondrotin sulfate component for gel solutions with different 
cross-link fractions.}\label{fig:Fig3}
\end{figure}

Fig. \ref{fig:Fig3}a,b,c depict the equilibrium volume fraction, corresponding Donnan pressure and the net-osmolarity, respectively, vs. the weight-fraction $\alpha$, of the charged component of the gel (i.e. ChS) and for different wt/vol percentages. The gel is dissolved in salt-free, neutral water ($[H]_b=10^{-7}$ M), and the negative charges on the gel causes water to dissociate into hydrogen and hydronium ions. H$^+$ ions can then bind with the monomers with the binding reaction
\beq
{\rm M^{2-}}+{\rm H^+} \xrightleftharpoons[\text{k$_{-h}$}]{\text{k$_h$}} {\rm MH^-} , \hspace{0.25cm} {\rm MH^{-}}+{\rm H^+} \xrightleftharpoons[\text{k$_{-h2}$}]{\text{k$_{h2}$}} {\rm MH_2} \label{eq:H}
\eeq
%

A chondrotin sulfate-dominant gel solution swells (i.e. $\theta_p$ decreases vs. $\alpha$, Fig. \ref{fig:Fig2}a), and has a higher osmolarity at equilibrium (i.e. $|$Osm$|$ increases vs. $\alpha$, Fig. \ref{fig:Fig2}b). This is because the swelling is driven by Donnan pressure (Fig. \ref{fig:Fig3}b) as well as the osmotic pressure (Fig. \ref{fig:Fig3}c) which is non-negligible in a charged gel. The negative charges on the gel causes water to furnish H$^+$ ions which contributes to a non-zero ionic difference (or `Net Osmolarity') across the gel. Similarly, gels with a higher cross-link fraction prefer a de-swelled equilibrium state at low ChS weight fractions and then exhibit a more prominent swelling at higher ChS weight fractions (for e.g., compare the swelling profile of the case CL$_3$ versus the case CL$_1$, in the range $\alpha < 0.25$, and $\alpha > 0.40$, respectively, Fig. \ref{fig:Fig3}a). Experiments corroborate that a highly cross-linked network provides a barrier against the Donnan pressure and hence acts against swelling \cite{Muir1970}.
 
%
%
%
%
%
%
\subsection{Effects of changes in the bath salt concentration} \label{subsec:SC}
\begin{figure}[htbp]
\centering
\subfigure[]{\includegraphics[scale=0.5]{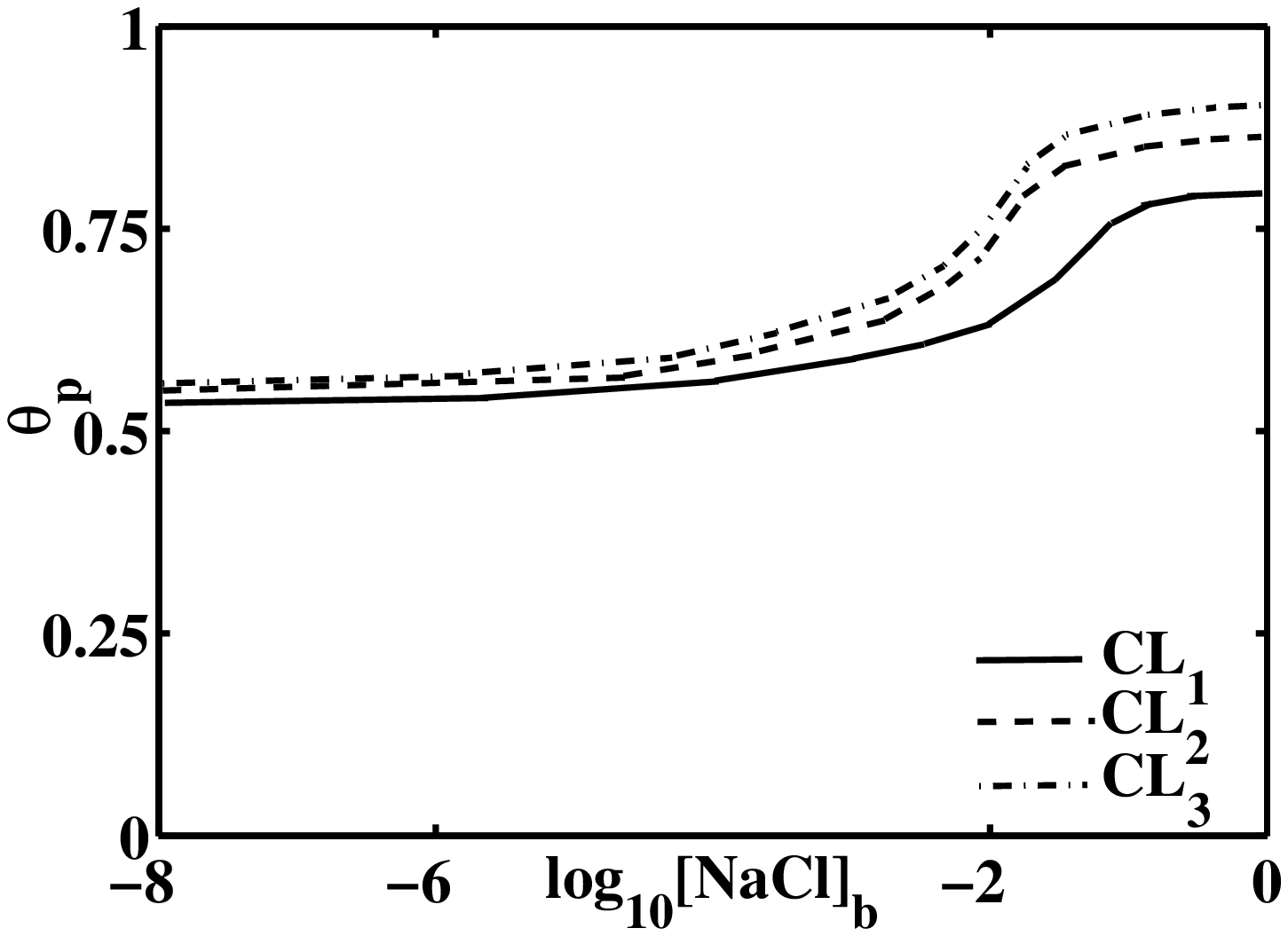}}
\subfigure[]{\includegraphics[scale=0.5]{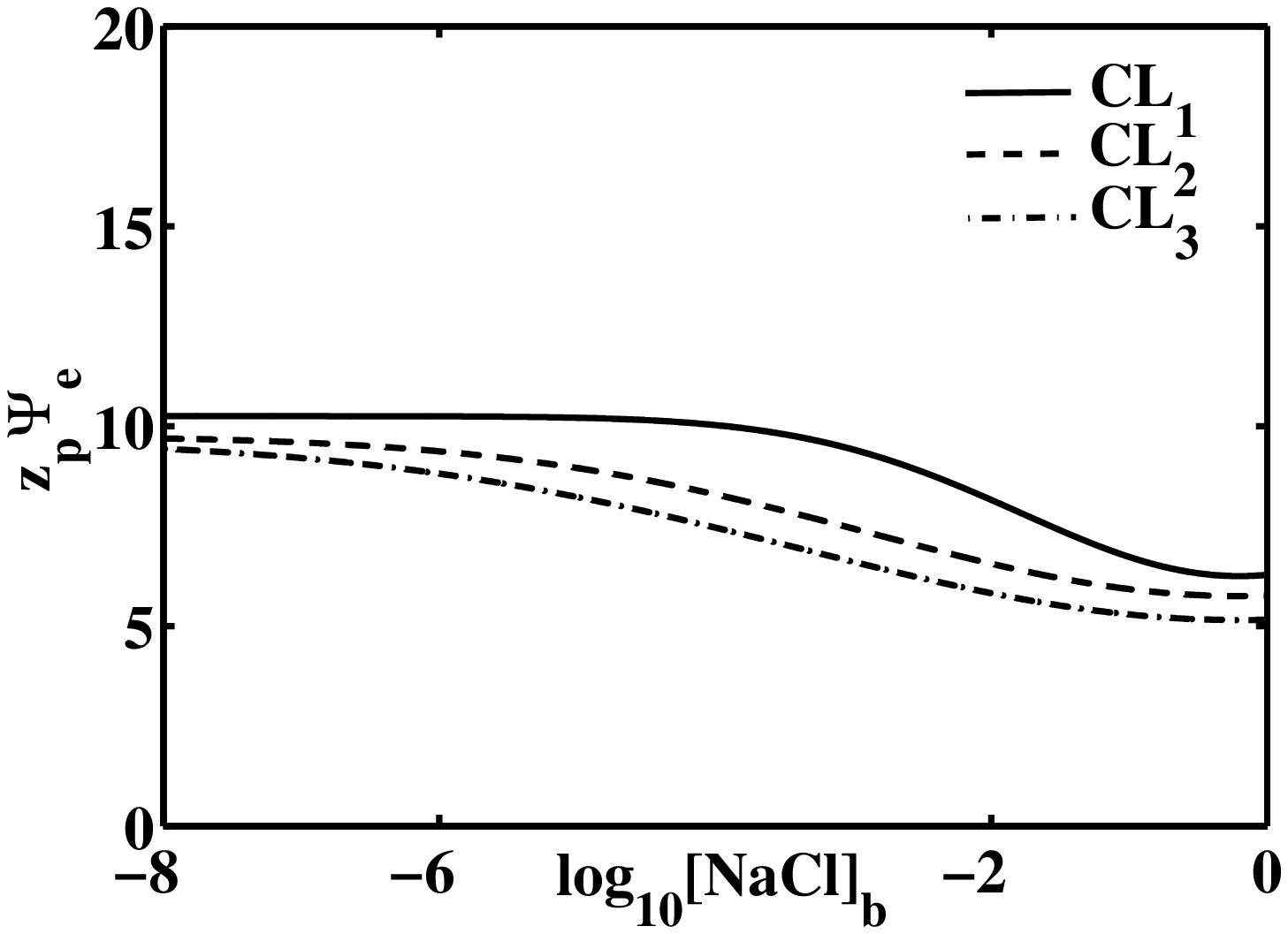}}
\vskip 0.00001cm
\subfigure[]{\includegraphics[scale=0.5]{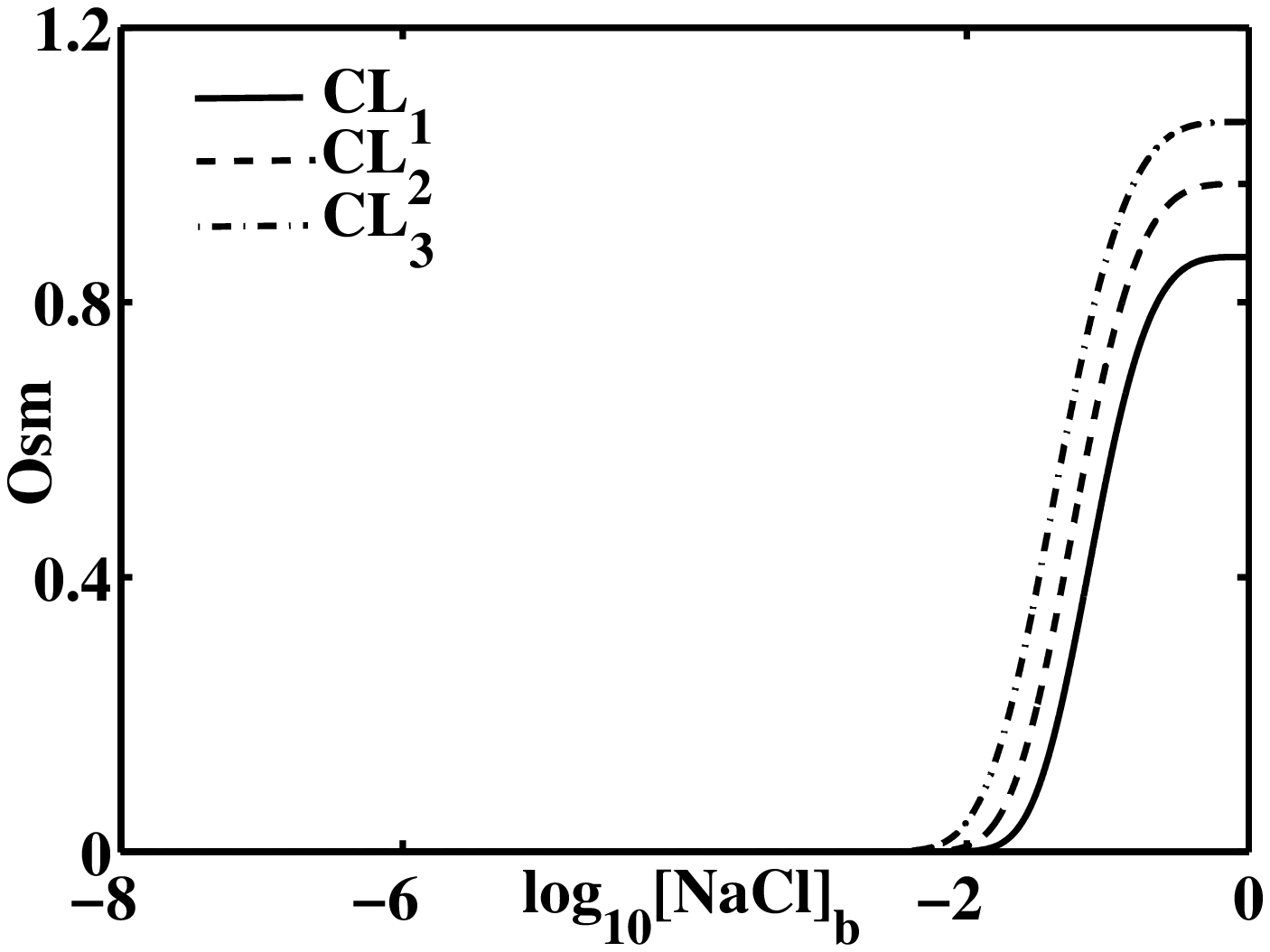}}
\caption{(a) Equilibrium polymer volume-fraction, (b) Donnan swelling pressure, and (c) Net-Osmolarity, vs. NaCl-concentration (in mol/lt.) in the bath. Different curves represent samples with a fixed 10\% ChS weight percent, but with a variable cross-link fraction.}\label{fig:Fig4}
\end{figure}

Next we explore the situation in which the gel is immersed in an infinite bath, with fixed hydrogen concentration ($H_b = 10^{-7}$ M), containing a monovalent ion, e.g., Na$^+$. Both hydrogen and sodium ions compete to bind with the negatively charged gel and the list of binding reactions are given in Eqn. (\ref{eqn:chem}). Fig. \ref{fig:Fig4}a,b,c highlight the equilibrium volume fraction, Donnan pressure and the net-osmolarity vs. the bath salt concentration, [NaCl]$_b$, respectively, for gels with different cross-link fractions but a fixed 10\% ChS composition (by weight) out of the total polymer weight. 

A higher salt concentration promotes gel de-swelling (e.g. consider volume fraction values at [NaCl]$_b$=1 M vs. those at [NaCl]$_b=10^{-3}$ M, Fig. \ref{fig:Fig4}a). A high salt concentration implies that more Na$^+$ ions are available to bind with the gel, thereby reducing the Donnan swelling pressure.
Although the osmotic pressure increases at high salt concentrations (Fig. \ref{fig:Fig4}c), it is the Donnan pressure that dominates the swelling mechanism in this case.
 
At lower salt concentrations, however, swelling is insignificant (i.e. $\theta_p \approx 0.55$ for concentrations [NaCl]$_b < 10^{-6}$ M). At these low concentrations water breaks down to produce H$^+$ ions to bind with the gel, which keeps the Donnan pressure relatively fixed (i.e. $z_p\Psi_e \approx 10$ for [NaCl]$_b < 10^{-6}$ M, Fig. \ref{fig:Fig4}b) and the osmotic pressure is nearly zero (Fig. \ref{fig:Fig4}c). Gels with higher cross-link fractions provide greater resistance against the Donnan pressure and therefore prefer a de-swelled state (i.e. compare the volume fraction values for the different curves in Fig. \ref{fig:Fig4}a). 

\begin{figure}[htbp]
\centering
\subfigure[]{\includegraphics[scale=0.5]{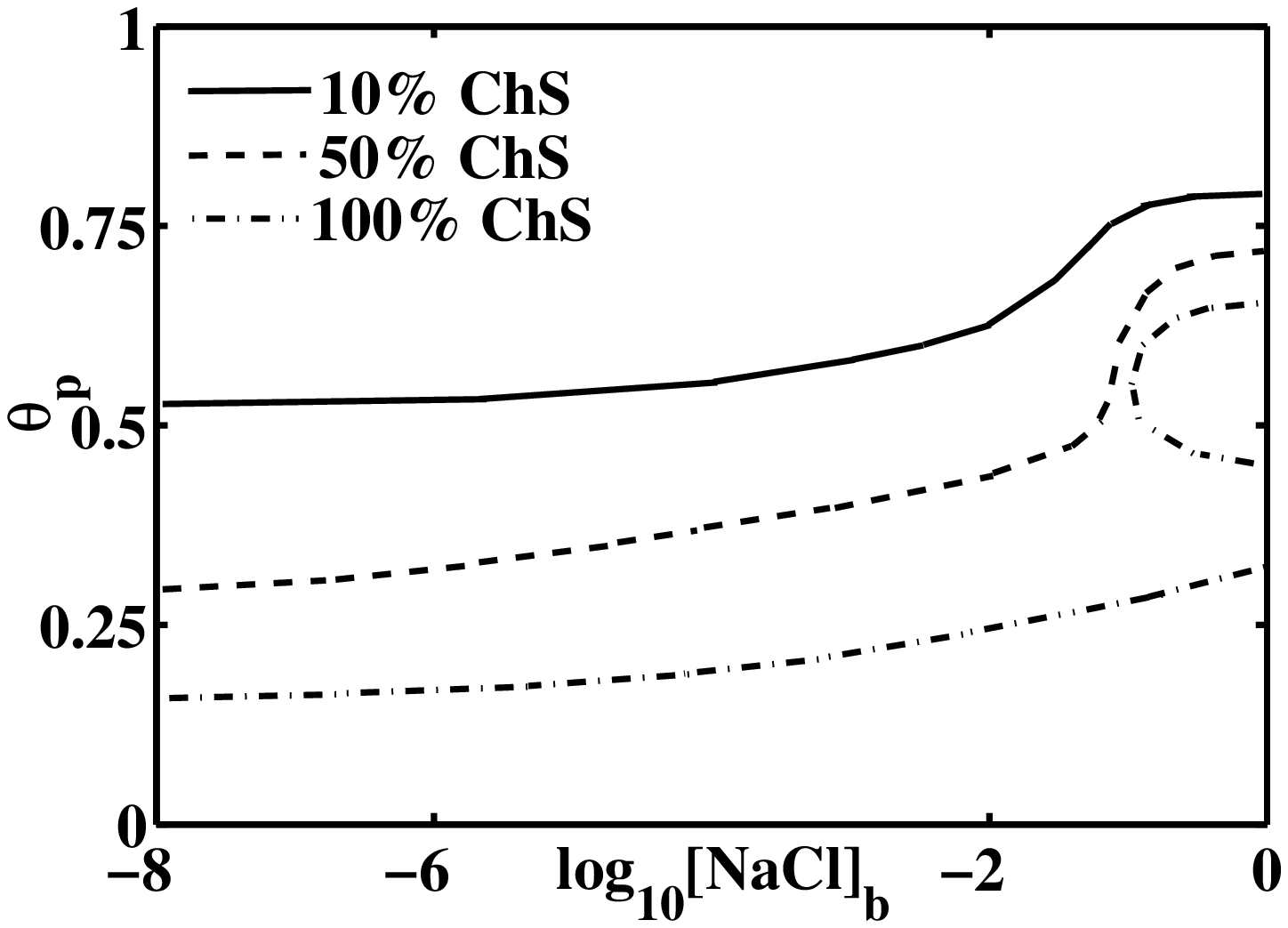}}
\subfigure[]{\includegraphics[scale=0.5]{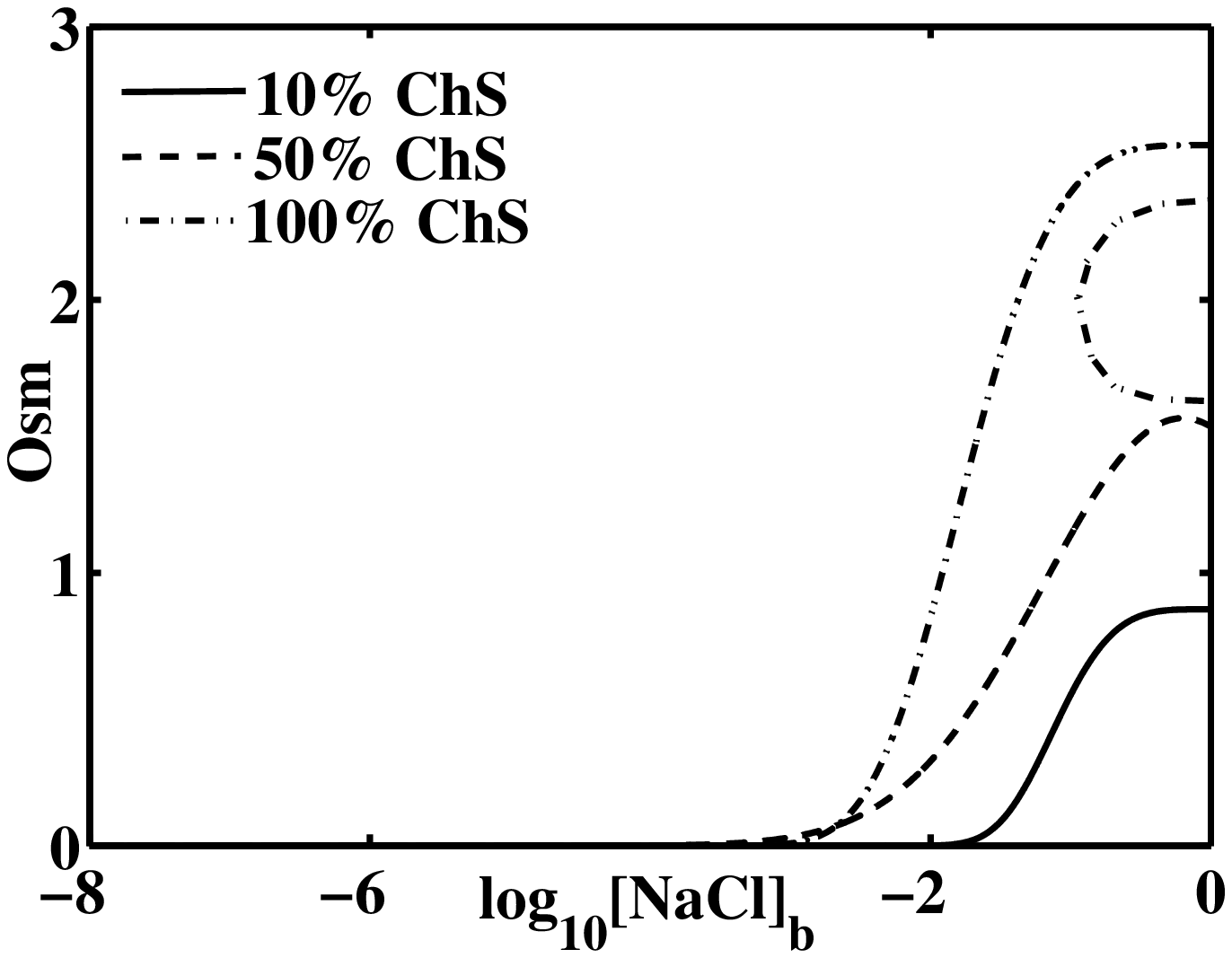}}
\caption{(a) Equilibrium polymer volume-fraction, and (b) Net-Osmolarity, vs. NaCl-concentration (in mol/lt.) in the bath. Different curves represent gel samples with a fixed cross-link fraction, $k_1=0.25$, $k_2=0.25$, and with a variable ChS weight percentage.}\label{fig:Fig5}
\end{figure}

Fig. \ref{fig:Fig5}a,b show the equilibrium swelling-deswelling state and the net-osmolarity vs. the bath salt concentration, respectively, for gels with variable ChS weight percentage but the cross-link fraction fixed at $k_1=0.25$, $k_2=0.25$. Once again, at high salt concentrations the gel de-swells (Fig. \ref{fig:Fig5}a). This is due to the lowering of average charge per monomer and hence the Donnan pressure, the driving mechanism for gel swelling. Comparing the different curves in Fig. \ref{fig:Fig5}a, we find that higher weight percentages of ChS result in higher average charge per monomer which leads to gel-swelling, driven by the Donnan pressure as well as the osmotic pressure (Fig. \ref{fig:Fig5}b).


\subsection{Effects of changes in the bath pH} \label{subsec:pH}

\begin{figure}[htbp]
\centering
\subfigure[]{\includegraphics[scale=0.5]{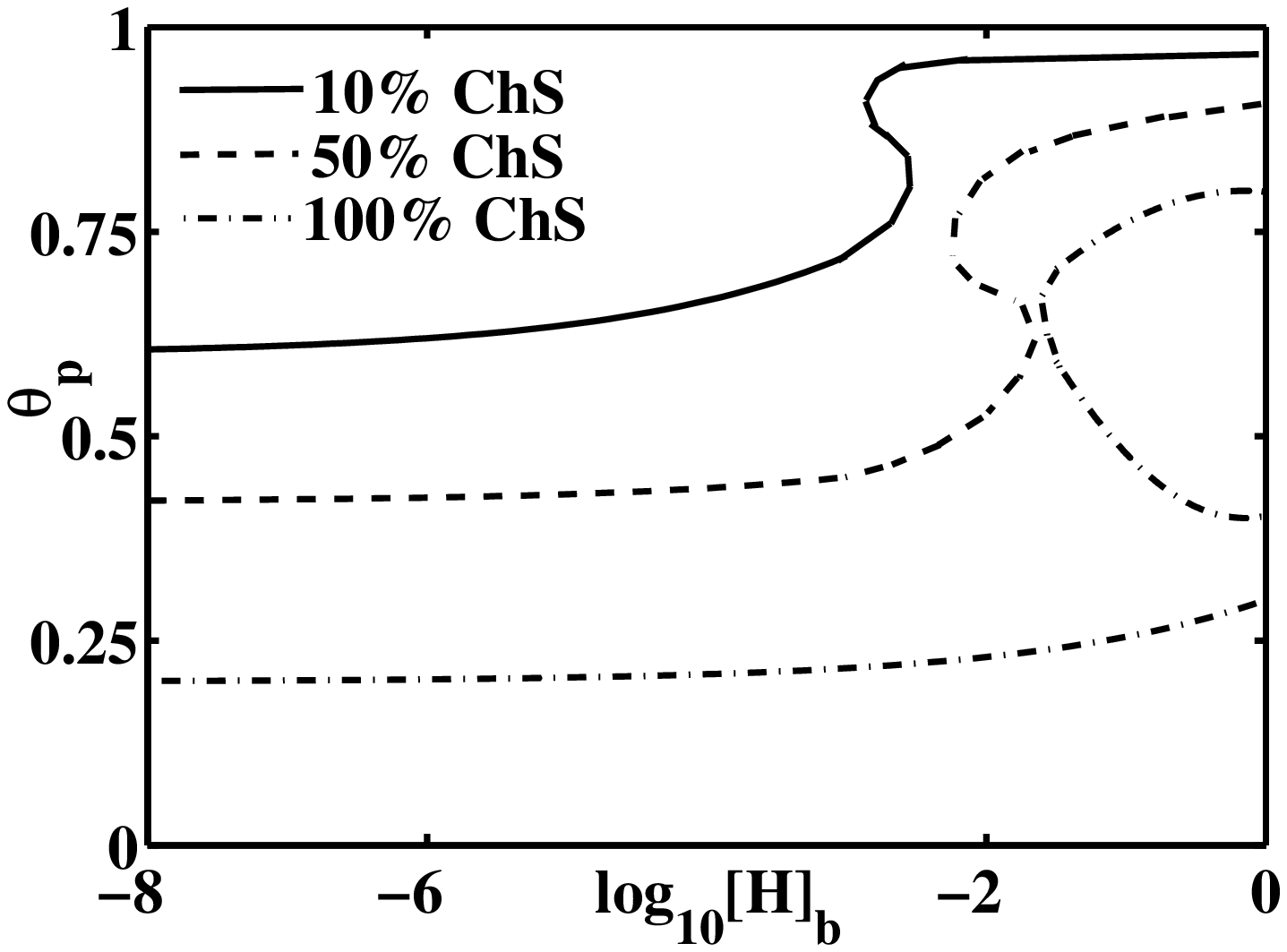}}
\subfigure[]{\includegraphics[scale=0.5]{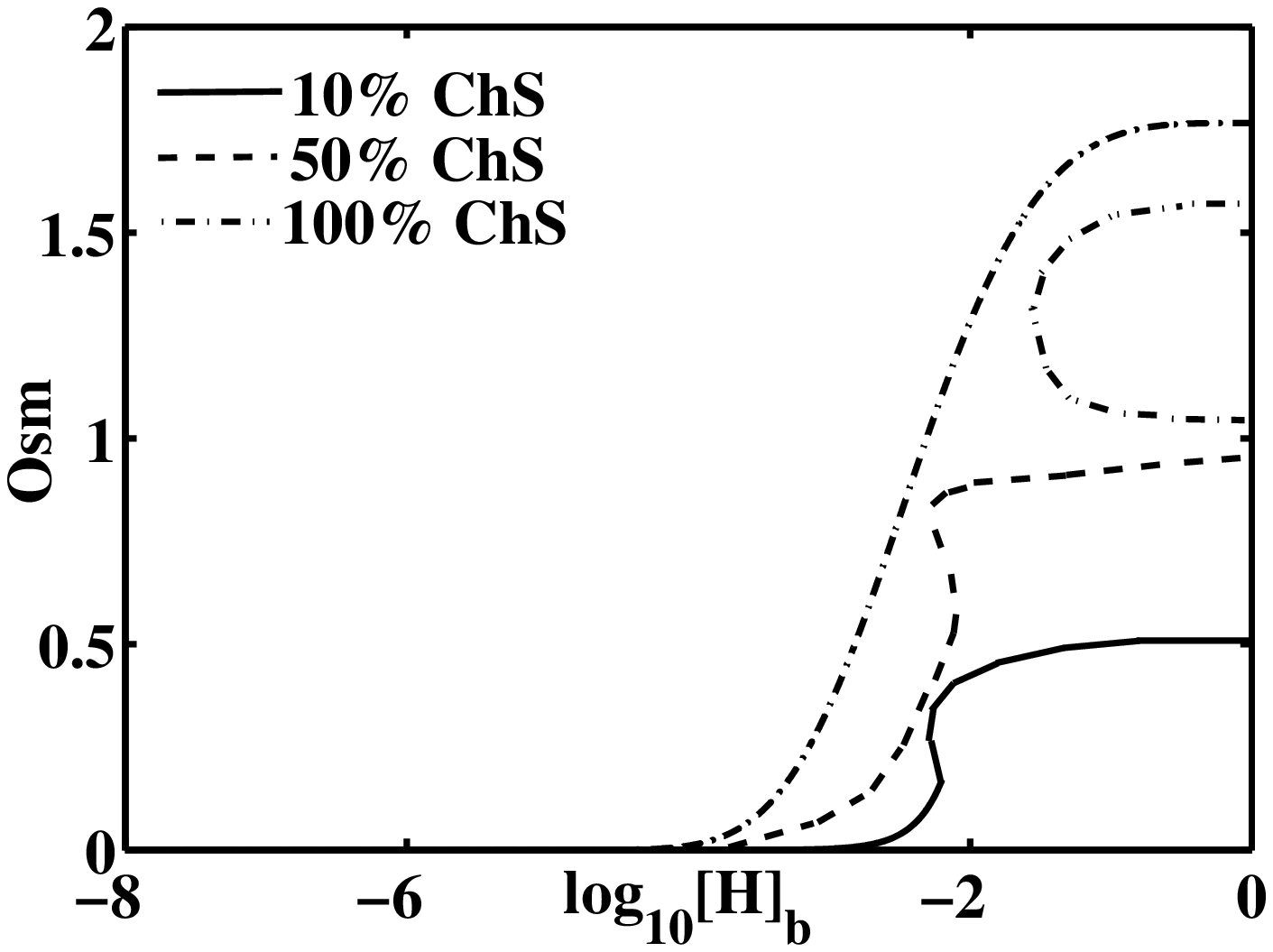}}
\caption{(a) Equilibrium polymer volume-fraction, and (b) Net-Osmolarity, vs. bath pH, for a fixed 10\% wt/vol gel.}\label{fig:Fig6}
\end{figure}

Finally, we predict the equilibrium configuration (Fig. \ref{fig:Fig6}a) and the net osmolarity (Fig. \ref{fig:Fig6}b) of gels (the cross-link fraction fixed at $k_1=0.25$, $k_2=0.25$ and variable ChS weight percent) immersed in salt-free, acidic/basic solution. The binding reactions of the charged polymer with the ions are those given in Eqn. (\ref{eq:H}). In a variable acidic conditions (e.g. log$_{10}$H$^+$ $> -1.9$), the gel de-swells sharply either via a continuous / reversible transition (i.e. referring to the changes in the 10\% and 50\% ChS-curves) or discontinuous / irreversible jump (for pure ChS polymers). These transitions highlight a non-linear swelling/de-swelling response mechanism of the ionic gels with respect to the pH variations in salt-free solution. 
%
%
%

\section{Conclusions} \label{sec:Conclusions}
In this paper, we have developed a multi-phase, multi-species model to quantify the swelling/de-swelling mechanism for polylelectrolyte gels. Using this model we have quantified the effects of the changes in bath concentration of monovalent solute (i.e. [NaCl]), the average charge per monomer (via variations in the chondrotin sulfate mass percentage) and the cross-link fraction of the gel on the equilibrium swelled/de-swelled configuration. We learned that, generally speaking, increasing the bath concentration of the ion species as well as the cross-link fraction of the gels per volume of total solution leads to deswelling while increasing the average charge per monomer leads to swelling, in agreement with experimental observations \cite{Bryant2003, Bryant2004, Bryant2005}. However, because of complex interactions between competing forces (e.g. Donnan forces which aids swelling and the elastic forces from covalent cross-links which helps de-swelling), the swelling/de-sweling mechanism is non-linear (or hysteretic) exhibiting a first order (or discontinuous) transition or a second order (or continuous) transition in many situations. A change in the solute concentration leads to changes in the equilibrium swelling of the gel which is either irreversible (in the case for first order transition) or reversible (for second order transitions). The full study of the kinetics of swelling these gels under spatio-temporally varying mechanical loads will be the subject of a forthcoming article.

\vskip 20pt

\noindent {\bf Acknowledgement}: This work was supported in part by grants NSF CAREER 0847390. The authors would also like to thank Dr. James P. Keener (Department of Mathematics,  University of Utah) for providing for providing useful insights at various stages of model development.



\end{document}